Random Group Problem-Based Learning in Engineering Dynamics

Luciano Fleischfresser

*lfle@utfpr.edu.br*

Universidade Tecnológica Federal do Paraná

*Câmpus* Campo Mourão

Abstract

Dynamics problem-solving is highly specific to the problem at hand and to develop the general mind framework to become an effective problem-solver requires ingenuity and creativity on top of a solid grounding on theoretical and conceptual knowledge. A blended approach with prototype demo, problem-based learning, and an opinion questionnaire was used during 1$^{st}$ semester of 2013. Students working in randomly selected teams had to interact with classmates while solving a randomly selected Dynamics problem. The approach helps improve students' awareness of what is important to learn in this class while reducing grading load on the instructor. It also provides a more rewarding contact time for both pupils and instructor.

Random Group Problem-Based Learning in Engineering Dynamics

**Introduction**

Devising a learning management system (LMS) for non-mechanical engineering majors taking Vector Dynamics is a challenging task. In this paper, one such system used during Fall of 2013 and coined Random Problem-Based Learning (RPBL) is introduced. This LMS blends aspects of traditional lecturing with active learning techniques in an attempt to get students to face upfront a major goal of a course like this: to learn how to solve problems of rigid body motion on paper.

Dynamics is the last part of a two-course sophomore sequence for many Engineering programs known as Vector Mechanics. The first part is a course on Statics where methods to solve the mechanical equilibrium of rigid bodies are developed. Statics is also a natural prerequisite for Mechanics (Strength) of Materials covering topics on internal stresses originating from external loads / thermal processes and the ensuing deformations from actions like torsion, buckling, bending and shear. The last two courses are very important core engineering prerequisites for Civil Engineering majors and these students tend to see such classes as important ones for their education. Dynamics may attract their interest on what it relates to the theory of vibrations for example, but not much more. On the other hand, Mechanical Engineering majors see it as highly important for their education, since much of the kinematics and force-torque considerations are applicable in later courses in their programs. It is marginally of interest to Electronic Engineering majors as it relates to robotics for example, but the content specifics of Dynamics are probably not all that attractive in this major's mind. This is not to mention Chemical, Environmental, and many other specializations that may have the full sequence of Mechanics (Statics, Dynamics and Strength of Materials) as required coursework to get a degree. These majors may not relate directly with these courses in view of what will be required of them when joining the workforce.

The LMS described here is a proposal to:

• deliver Dynamics problem-solving skills for non-mechanical engineering majors;

• promote active, peer-group learning with room for different learning styles;

• reduce grading workload, a major bottleneck for mid to large enrollment classes;

• improve contact time for all.

**Background**

Not all schools have Dynamics in their offerings of core engineering classes. The experience gained as a faculty member in a school that offers it to all engineering majors makes one revise its approach to be more in tune with modern audiences. Students may ask: *why do we have to study this subject since it is not even close to what I will do after graduation*? Or they might reason: *since this is a required class, let's work our way just to make ends meet at the end of the semester*. Of course these are just guesses in a professor's mind, but students' evaluations at the end of the semester give some clues that these thoughts may, in fact, surface at times.

An effective tool to assess students' preparedness is the Mechanics Readiness Test, a 25 multiple-choice exam that assesses the background knowledge and skills of students at the beginning of a Dynamics course (Snyder and Meriam, 1978). This instrument was used with four different sections during the first week of class in 2012. The outcome is shown in figure 1 with bars separating results by section. The only majors participating were Civil, Electronic, Environmental and Food Engineering. A Mechanical Engineering program of study is not offered. While students may enroll in any given section, there are preferences. Students from a major other than the preference for a given section may register if seats are still available after main enrollment is completed. Section 1 is given preference for Environmental Engineering, section 2 for Food Engineering, section 3 for Electronic Engineering and section 4 for Civil Engineering. The results uncovered that students enter Dynamics with poor background knowledge and skills needed to perform well. It should be noted that the majority of these students already took a Statics course, and they may be taking Strength of Materials and Fluid Mechanics/Heat Transfer concurrently. These classes rely heavily on freshman mathematics and physics. Grading of math and physics related undergraduate work is not a straightforward task if the instructor is willing to assess worked-out solutions instead of final answers only. Moreover, these are usually medium to large-enrollment classes (30+ students) and traditional lecturing alone is widely regarded as an inferior mode of instruction for the Internet generation (Belytschko et al., 1997), (Rutz et al., 2003). Approaches based on rigorous educational research are being developed and applied at an ever-increasing rate (Menekse, Stump, Krause, Chi, 2013). Methods like problem-based learning (PBL) (Qin, Johnson, Johnson, 1995) are suitable for the subject matter of Dynamics. Goodhew (2011) introduces a shared collection of concept questions that might prove useful as a resource available for consultation (in an on-line platform for example).

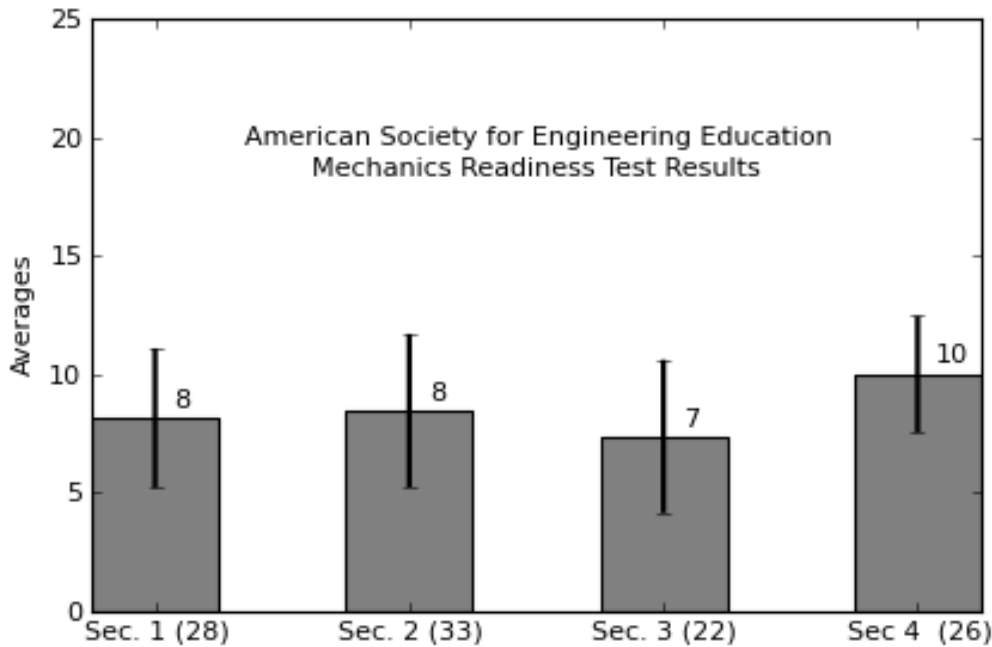

**Figure 1** . Results for the Mechanics Readiness Test among various Dynamics sections during the 2012 academic year. Bar heights are the averages among all students taking the test for that particular section. Black lines are standard deviations. Number of students taking the test in each section is in parenthesis under horizontal axis. See text for details.

In view of these considerations, a revamped offering of this course is being developed over the years to find answers to the following questions:

1)   How can one effect change in the classroom environment from one where students behave as passive listeners and note takers to one where they engage with classmates to discuss and construct problems' solutions as a team?

2)   What elements of traditional lecturing are enduring ones, and what elements should be deemphasized to optimize learning for all students while delivering key concepts and skills on the subject matter of Dynamics?

## Method

With these issues in mind, a LMS built around a problem-based learning (PBL) approach is proposed. Class time is spent in problem-solving sessions and discovery-based activities involving the use of a physical model. These are the traditional parts of the course where the instructor stands in front of the classroom and uses the chalkboard to write sample problems' solutions , explain some theory, and relate the content with a physical model built to aid teaching.  A student poll is used to rate exercises in each problem set as easy, intermediate or hard. The use of a student poll releases some of the control a professor has on grading, empowering students with a sense of ownership over their assessment, along the lines of self-determination theory (Niemiec and Ryan, 2009). A feature of the approach is a random selection of 3-4 students to form teams to solve a randomly selected problem during examinations. Student teams are also required to select problems to present orally in front of their classmates at the end of the semester. These are the PBL components of the course. On exam day , desks and chairs are re-arranged as "learning centers" with students facing each other in their groups and working together to prepare written solutions for the ramdomly selected problems.

### *Random Grouping – The PBL part*

Class material was roughly divided in 1) kinematics of rigid bodies; 2) rotational inertia and kinetics of general plane motion and; 3) work-energy and impulse-momentum methods for rigid bodies.  While the exam was taking place, team behavior was observed and annotated. After tabulating results, the midterm examination took into account the information collected during the first test, and redistribution of problems attempted to leverage difficulty based on that information. Midterm teams were formed using a *selective* random draw taking into account these considerations. It was possible to have recurring members in a given team but an effort was made to leverage expertise among students, placing the more advanced solvers in separate groups. The third problem set wrapped up the material involving application of the accumulated theory and practice built throughout the course. Teams were once again redistributed randomly as in the first test.

    Full solutions to all problems were made available after the tests. Students would only have the answer key as a guide beforehand. Since they would not know in advance which problem they would receive on exam day, the more conscious students would attempt all problems in preparation for the test. Risk takers would rely on teammates during exam, but they would not know who would be in their teams until exam day. The approach encourages peer cooperative learning. Although no learning style survey was undertaken, the random selection of groups naturally reaches students with different learning styles and individual skills.

    To give an idea on how teamwork took place during exams, a fictitious class with 16 students is used as an example. Here, 4 teams with 4 students each can be formed (letter names):

                  A-B-C-D;     E-F-G-H;     I-J-K-L;     M-N-O-P

In this example, a problem-set is minimally composed with 4 exercises, but it could be more. For the first problem-set, one easy, two intermediate and one hard could be selected. The teams for the first assessment could be as follows:

A-B-C-D → intermediate;    I-J-K-L → hard;

E-F-G-H → easy    M-N-O-P → intermediate

For the 2nd exam, the following is possible:

A-B-C-D can get easy, intermediate or hard;    I-J-K-L can get easy or intermediate;

E-F-G-H can get intermediate or hard;    M-N-O-P can get easy, intermediate or hard

In summary, all 16 students can get an intermediate problem during the 2nd assessment, 12 students can get a hard one, and 12 students can get an easy one. Choosing the 2nd problem-set with one easy, one intermediate, and two hard problems, one

possibility for teams could be:

A-I-K-O → easy;    B-C-D-F → hard;

E-G-J-L → intermediate;    H-M-N-P → hard

With two assessments concluded, here are the possibilities for each student during exam 3:

A → hard;    B → easy;    C → easy;    D → easy ;

E → hard;    F → intermediate;    G → hard;    H → intermediate;

I → intermediate;    J → easy;    M → easy;    N → easy;

K → intermediate;    L → easy; O → hard;    P → easy

This arrangement could be realized with a minimal problem-set (4 exercises), being 2 easy ones, 1 intermediate, and 1 hard. The final tally is presented in table 2. Notice that, for this particular example, a perfect balance of problem difficulty is achieved for each student. That is, everyone would get one problem for each perceived difficulty using only the minimal quantity of exercises per problem set. In an actual setting this fair combination may not be minimally possible. In those cases, the instructor needs to offer more exercises per problem-set to attain perceived fairness.

| Student | Problem Set 1 | Problem Set 2 | Problem Set 3 |
|---|---|---|---|
| A | intermediate | easy | hard |
| B | Intermediate | Hard | Easy |
| C | Intermediate | Hard | Easy |
| D | Intermediate | Hard | Easy |
| E | Easy | Intermediate | Hard |
| F | Easy | Hard | Intermediate |
| G | Easy | Intermediate | Hard |
| H | Easy | Hard | Intermediate |
| I | Hard | Easy | Intermediate |
| J | Hard | Intermediate | Easy |
| K | Hard | Easy | Intermediate |
| L | Hard | Intermediate | Easy |
| M | Intermediate | Hard | Easy |
| N | Intermediate | Hard | Easy |
| O | Intermediate | Easy | Hard |
| P | Intermediate | Hard | Easy |

**Table 2.** Final distribution of problems per student for full semester.

*Physical Model – The traditional part*

The use of a physical prototype is discussed next. The figures for the solved problem (Shames, 1997) and for the physical model are shown below (figures 2 and 3). The problem asks for linear velocities' instantaneous values for points A and B at position shown, and angular velocities and accelerations for bars AB and BC. A follow-up exercise is to find the instantaneous center of rotation for bar AB. The physical model helps illustrate the idea of a fixed vector in the rigid body, visualization of the full motion of the mechanism, the individual motions of each part, and common points of two given rigid bodies (connection pins). A few concepts commonly misunderstood are the difference between linear and angular velocities and accelerations, the vector nature of rotational quantities, and the proper use of Chasles theorem to solve the kinematics.

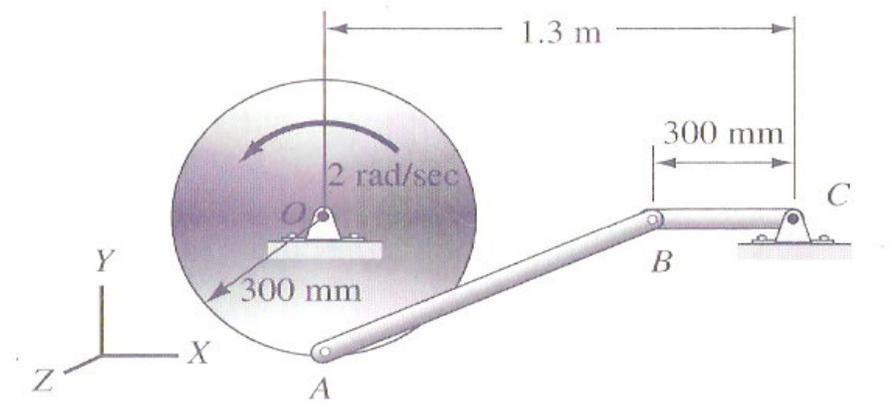

Figure 15.17. Two-dimensional device.

**Figure 2.** Textbook figure of Shames problem.

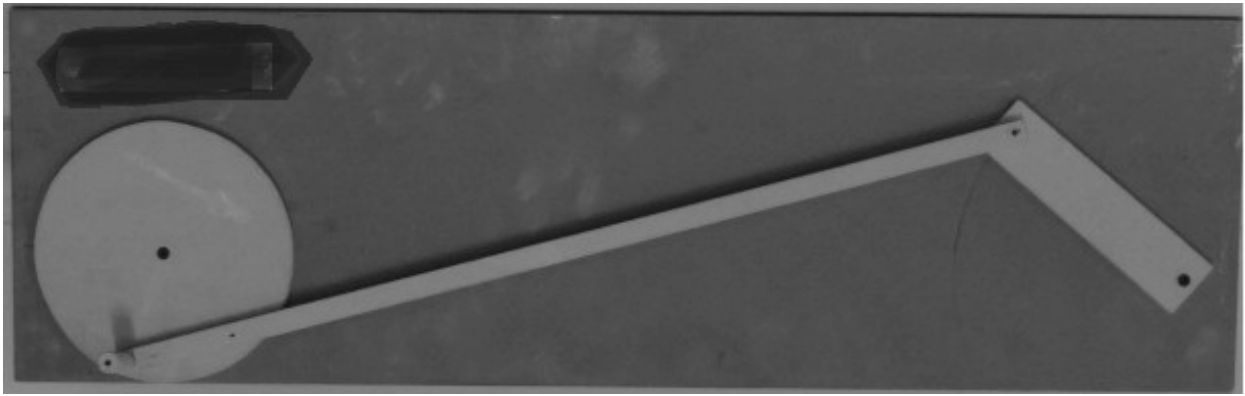

**Figure 3.** Physical Model to illustrate kinematics of rigid bodies.

**Discussion**

After concluding the semester it became apparent that there were mixed feelings about the adopted procedure. Tabulating historical approval rates for all Dynamics offerings since 2008, one sees that 2013-1 was the semester with the lowest approval rate closely followed by 2010-2 when active learning was adopted utilizing weekly quizzes, a time-consuming approach for both students and instructor (figure 4). Selected students' comments are presented below. These are anonymous remarks usually recorded after grades are in:

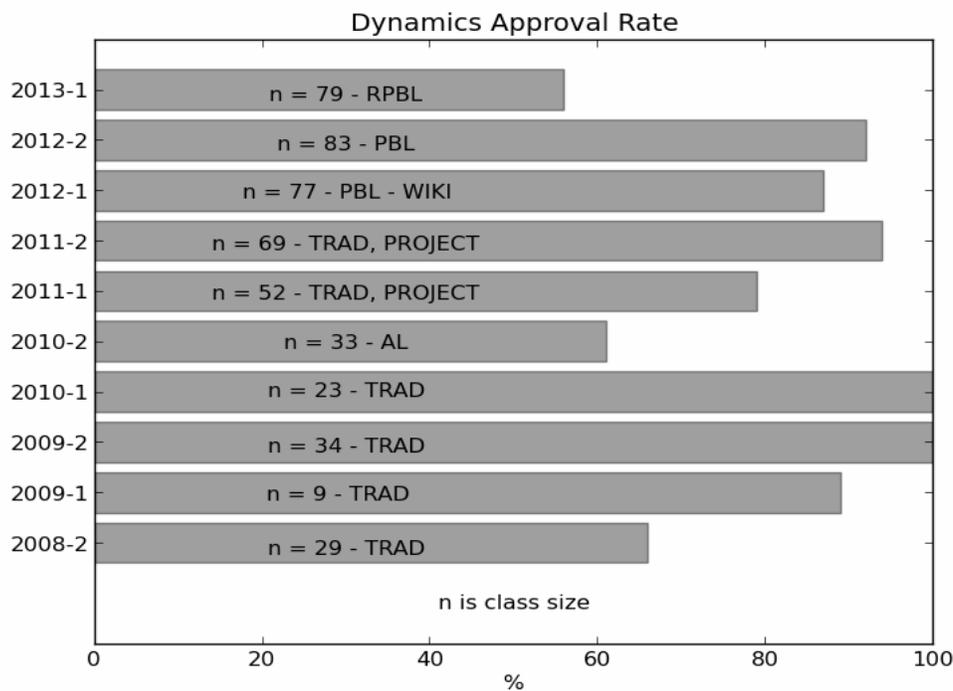

**Figure 4.** Historical approval rates since initial offering of Dynamics. The meaning of the descriptors are: TRAD: traditional lecturing; AL: active learning; TRAD, PROJECT: traditional lecturing with semester-long computer project; PBL-WIKI: Problem-based learning with a wiki platform for students to submit their semester-long project (article summary); PBL: problem-based learning without exams, grade accumulated with various problems handed by groups of students; RPBL: random problem-based learning being described in this work.

"Unfair assessment".

"The professor could change the assessment method. Comparing the exercises from the 1st problem set with the ones from the last problem set, one sees major differences in level of difficulty. The students that should had received easy problems in last assessment, ended up being sacrificed".

"In the beginning the assessment method was somewhat confusing. I believe that, even though the proposed method helps the student a lot (in general), one still has to rely on luck. A student in the average level can have the misfortune of getting ONLY ONE exercise that he/she will not be able to solve".

"A theory is needed before explaining the exercises and it has to be explained clearly and in detail. It is important to explain the content and provide problem sets for the students to try. He should use books where we can find *[sic]* "on the Internet", and not books where neither solutions nor similar examples can be found."

"The learning system is completely wrong as well as the assessment method where the exercises are literally randomly selected. In any given exam, one student can get a simple integral exercise while another can get a problem that requires two pages to be solved. This makes the process totally unfair and it does not assess students' knowledge".

Overall, comments are not favorable and they expose a general feeling of unfairness with the assessment performed. They also expose misunderstanding with what is being attempted. For example, the second comment does not take into consideration that it is expected that students progress on their own learning curves as the semester unfolds. So it is natural that the average difficulty of problems at the end is somewhat higher than at the beginning of the semester. A perceived easy problem at the end, if used in the beginning, might had been chosen as an average or difficult one. The third and last comments ignore the fact that *teams of* students work on a given problem, and not a single student. The last comment also overlooks the fact that leveraging is being sought throughout the semester and not on a single assessment. For the most part, students interacted in a productive manner during group problem solving. But their comments are prompting a re-examination of team selection. The current belief is that the random selection should be kept for the first exam, but adapted as the semester evolves and the instructor gains familiarity with students' skills. Team selection should be influenced by the instructor on the second exam with the goal of joining pupils at varied skill levels. At the end, only minor adjustments would be required for the third and final exam to optimize learning.

The results shown in figure 4 present teaching/learning methods utilized prior to the RPBL approach used in 2013-1. For example, 2012-1 and 2012-2 used a LMS with individual grade tracking and display in graphical format using password protected individual web pages (figure 5). There were no formal exams, but problems were offered for students to solve in-class, or to take home and bring next class. Weights for each activity were specified in the beginning of the semester to accumulate a 10-point total at the end. The major difficulty with this approach - from an instructor's point of view - was that one would have to be continuously updating grades for each and every student. This became an almost daily task during the given semester, and it required a backlog of all students' records. Project-based learning was introduced during 2011-1 and 2011-2. These were computer projects extending class examples where an initial computer code was provided and student teams would have to modify it to solve a dynamics problem. Project assignments were semester-long activities and, overall, they were good learning experiences for students. Comparisons between problem-based and project-based learning among three college majors (2 Engineering and Medicine) revealed that the best predictors for student engagement and persistence are stress-related (Bédard et al., 2012). Factors like autonomy, support, and flexibility in the curriculum are important for students to persist and to engage.

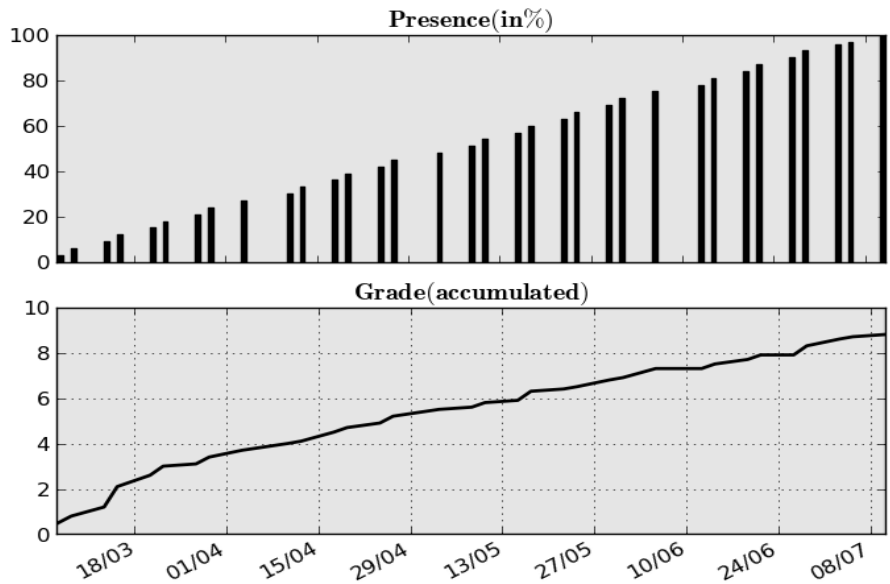

**Figure 5.** Sample of graphical display used during 2012-1 and 2.

It is apparent that others have struggled with and attempted to improve their teaching approaches for Vector Dynamics. For example, Gray and Constanzo (2008) coined the term PCA (Problem-Centered Approach) and reported an increase of students' interest in Dynamics. A group at Northwestern University (Belytschko et al., 1997) revamped their offerings of introductory core engineering classes to an Engineering First mode with emphasis on computer solutions for Mechanics problems. They reported on improved outcome and better motivation of students taking Mechanics. The idea of bringing tangible examples to the classroom like the physical prototype of figure 4 helps guide inquiry during class discussions. It has been explored in the context of stoichiometry for engineering students by Pinto and Prolongo (2013). There too, the authors report on improved student motivation and critical thinking skills. According to Kipper and Rüütmann (2013), instructors need to have knowledge in at least four different areas for expert teaching: knowledge of content; pedagogical content knowledge; general pedagogical knowledge; and knowledge of learners and learning. They state that achieving this teaching level helps students understand the material and develop good critical thinking skill habits. An interesting approach to understand students' problem-solving skills is to observe eye gaze patterns while reading problem information (Smith, Mestre and Ross, 2010). It has been reported that students actually retain very little of the conceptual information available in the problem question despite spending almost half of the time reading the text contained therein. Spatial ability has also been reported as influential in kinematics problem-solving (Kozhevnikov, Motes and Hegarty, 2007).

An important conclusion from self-determination theory (SDT) is that the use of controlling strategies in instruction has a detrimental effect on analytical problems' reasoning when compared with non-controlling strategies (Boggiano et al., 1993). Surprisingly, students perceive the instructor that imposes more control as being a better teacher despite their poorer performance. Extrinsic motivation is one important factor of SDT, and students' prior experiences may interfere with instructor's efforts to implement the approach in the classroom.

A much broader view on students' motivation is provided in Jones et al. (2013). The discussion is geared towards capstone design engineering courses, but the MUSIC model is quite general and easily adaptable to an introductory Engineering class like Dynamics. Furthermore, it helps discern the easiness of implementation in a given situation. Interest and usefulness are probably the two critical components to maintain in Engineering Dynamics for non-Mechanical Engineering majors. More specifically, individual interest and usefulness are hard to convey as students evaluate course content and their future aspirations as Environmental or Food Engineers for example. The other three components (empowerment, success, and caring) rely more on a professor's own attitude toward students and the material to be learned.

**Conclusions and Further Work**

A random problem-based learning (RPBL) method was introduced to improve learning of Dynamics for engineering sophomores. The general goal was to move away from a strictly traditional lecturing environment to one where students have more autonomy during class time and acquire a stronger sense of relatedness with classmates, the instructor and the course. The RPBL proposed here reduced grading workload while allowing one to focus on learning maximization.

The initial goals of the proposed RPBL were to deliver problem-solving skills for non-Mechanical Engineering majors, to promote active learning, to reduce grading workload, and to improve contact time. We describe now to what extent these initial goals were met. The focus on solving various problems on the chalkboard provided students with plenty of opportunities to observe techniques employed by a more experienced person in dealing with the mathematics, the physics, and the spatial visualization issues related to the subject matter of Dynamics. Often, the formal delivery of the concepts and theory prior to application is a major deterrent of student learning and a major source of dissatisfaction for them. The learning center set up exposes students to one another, letting them compare their own comprehension of topics and promoting leveraging of student thinking on the subject matter. Passive learners gain a better appreciation of the material and the role of the instructor. Group work certainly reduced the amount of grading as annotating 3 or 4 similarly presented solutions is much more time-efficient than individually presented ones. This issue becomes even more critical for mid-to-large enrollment classes in view of the need for fast feedback on students' performance. Finally, quality contact time is important to promote effective learning in the classroom environment. The elements being proposed in the RPBL approach are intended to attain this level of interaction. As mentioned in Svinicki (2005), achieving nirvana in this regard is unrealistic, but course planning should always aim for such a goal. In particular (and based on students' evaluations) it became clear a certain level of discomfort with the random selection of groups. Students prefer to choose whom to work with instead of being chosen. However, their motivations to form the groups usually do not match the instructor's best intentions, and so a compromise must be found. This is still an issue to be considered in future sections of the course. Taking into account these are sophomore students, the majority of them still don't have a clear picture of professional interests.

Many of the issues uncovered with the proposed RPBL method can be framed in terms of a PBLE (Problem-based learning environment) (Jonassen, 2011). There are components and cognitive scaffolds to be matched to learners' needs in solving different kinds of problems. In Engineering Dynamics, one can easily encounter problems of various types. For example, troubleshooting problems arise when proposing a solution upfront and asking students to verify its correctness and/or plausibility. Decision-making problems arise in obtaining a solution to a textbook problem that is known to be possible using alternate routes. Assessing students' solutions in this type of problem can help the instructor place pupils on their developmental trajectories. This, in turn, contributes to instructional design by aligning learning objectives with desired outcomes.

Further work is needed regarding assessment. The overall message from students' comments hints that better assessment methods would be beneficial and they would reduce attrition rates. A better register is needed for this purpose. Perhaps it does not need to be strictly based on written work, but it may need to add an interview element. Given the potential subjectivity of grading students based not solely on documented work, a better tool is needed to more accurately gauge individual performance.

   Some unanswered questions remain and they have implications for further research. For example, how can one account for individual differences in teamwork? The RPBL method attempts to emphasize leveraging among students without careful individual assessment. This might be related to the perceived feeling of unfairness that some students expressed in the evaluation survey at the end of the semester. Differentiated learning apparently contradicts leveraging. This is a limitation of the current work and future developments should focus on reconciling teamwork in PBL environments with differentiated student learning.

RANDOM PBL IN ENGINEERING        22Curso de Engenharia de Producao. (n.d.). Retrieved April 17, 2014, from Universidade Estadual Paulista "Júlio de Mesquita Filho" Campus de Bauru website, http://antigo.feb.unesp.br/dta/graduacao/curriculos/4402.pdf

Department of Biomedical Engineering. (n.d.). Retrieved April 17, 2014, from Peking University, College of Engineering website, http://bme.pku.edu.cn/en/contents/Undergraduate_136/773.html

Department of Energy Resources Engineering. (n.d.). Retrieved April 17, 2014, from Seoul National University website, http://ere.snu.ac.kr/en/university/universitySub01.asp

Goodhew, P. (2011). Concept Questions in Engineering: The Beginnings of a Shared Collection. Paper presented at the 7th International CDIO Conference. Copenhagen, Denmark, June 20-23.

Gray, G. L., Constanzo, F. (2008). A Problem-Centered Approach to Dynamics. Paper presented at the 2008 American Society for Engineering Education (ASEE) Annual Conference and Exposition. Pittsburgh, PA, June 22-25.

Jonassen, D. (2011). Supporting Problem Solving in PBL. *Interdisciplinary Journal of Problem-based Learning, 5*(2). Available at: http://dx.doi.org/10.7771/1541-5015.1256

Jones, B. D. , Epler, C. M. , Mokri, P. , Bryant, L. H. , Paretti, M. C. (2013). The Effects of a Collaborative Problem-based Learning Experience on Students' Motivation in Engineering Capstone Courses. *Interdisciplinary Journal of Problem-based Learning, 7*(2). Available at: http://dx.doi.org/10.7771/1541-5015.1344